\documentclass[
    ,final            
  ]
  {aipproc}

\layoutstyle{6x9}

\begin{document}

\title{Spectral Properties of Interstellar Turbulence via Velocity Channel
Analysis}


\author{Dmitri Pogosyan}{
  address={Physics Department, University of Alberta, Edmonton, AB, T6G 2J1,
Canada}
}

\author{Alex Lazarian}{
  address={Department of Astronomy, University of Wisconsin, Madison,  WI 53706,
USA} }

\begin{abstract}
In this presentation we review the link between the statistics
of intensity fluctuations in spectral line data cubes with underlying
statistical properties of turbulence in the interstellar medium.
Both the formalism of Velocity Channel Analysis for optically thin lines
and its extension to the lines with self-absorption is described.
We demonstrate that by observing optically thin lines from cold gas
in sufficiently narrow (thin) velocity channels one may recover
the scaling of the stochastic velocities from turbulent cascade,
in particular, Kolmogorov velocities give $K^{-2.7}$ contribution
to the intensity power spectrum.
Synthetically increasing the channel thickness separates out
the underlying density inhomogeneities of the gas.
Effects of self absorption, on the other hand, retain the velocity
signature even for integrated lines. As a result, intensity
fluctuations tend to show universal but featureless scaling of the power
$\propto K^{-3}$ over the range of scales.
\end{abstract}

\maketitle


\section{Introduction}

There is little doubt that interstellar medium is turbulent
(see reviews \cite{Arm1995,Laz99a,LPE02}).
Turbulence proved to be ubiquitous in molecular clouds \cite{Dickman85},
diffuse ionized \cite{Cordes99} and neutral \cite{Laz99b} media.
Recent years were marked by a
substantial progress in theoretical and numerical description of both
incompressible and compressible Magneto-Hydrodynamical (MHD) turbulent
cascade \cite{GoldSrid95,LG01,CL02a,CLV02a,CL03a,CL03b}.
MHD turbulence controls
many essential astrophysical processes including star formation, transport
and acceleration of cosmic rays, of heat and mass
(see \cite{Schli99,VOPGS00,NM01,CLV03,CLV02a}).

The progress in testing and advancing our theoretical understanding
of these astrophysical processes rely on recovering the actual parameters
of turbulent cascade from observational data.

For this, the statistical approach  is
useful (see \cite{MY75, Dickey95}). In fact, the turbulent medium
is, generally, a non-uniform distribution of matter with stochastic, random,
density $\rho({\bf x})$, that moves according to stochastic velocity field
${\bf u}({\bf x})$. Recovering the scaling of statistical descriptors
for stochastic density and velocity will allow us to get insight
into mechanisms,
responsible for turbulent driving. Among the two, the most direct information
about the turbulence is contained in the statistics of the velocity field,
while the statistics of density fluctuations usually provides
complimentary indirect way of testing turbulence.

A wealth of observational information is contained in spectral or
channel maps of the emission lines produced by gas tracing the
turbulent cascade. Both inhomogeneous distribution of the emitting
matter and its motions result in small scale fluctuations of the
emission intensity observed at a given velocity. The problem is to
relate the two dimensional statistics of intensity maps to
underlying three dimensional properties of the turbulence and
disentangle the effects of random motions from the effects of
density fluctuations. This problem is addressed in the velocity
channel analysis (VCA) \cite{LP00,LP04} which we review here. VCA
provides theoretical framework for quantitative interpretation of
observational data in terms of turbulent properties.

The observational data has been obtained for the variety of lines,
mapping different regions of our galaxy (or its satellites) and
different physical conditions - $21\;cm$ diffused $HI$ in outer
parts of the Galactic disk \cite{Green93}, towards the Galactic
center \cite{DMcSGG01} and in Small Magellanic Cloud \cite{SL01},
$CO$ lines in molecular clouds \cite{SBHOZ98}, $H_{\alpha}$ lines
in Reynolds layer \cite{MS96}. In all the cases the reported index
of the power spectrum of turbulence is close to $n \approx -2.7$.
This is really remarkable given that these results are obtained
for quite different observables and that $n=-2.7$ does not
correspond to Kolmogorov \cite{Kolm41} cascade, the one case when
some universality may be expected. To complicate the matter, some
lines are optically thin, given view for the full depth of the
emitters along the line of sight, while for the other (definitely
for $CO$, but also for $HI$ viewed towards the Galactic center)
self absorption is important. Here, using the formalism of VCA, we
will demonstrate how the universal power spectral scaling in the
data may be explained.

\section{Statistical description of the Turbulence}

\subsection{Stochastic density and velocity fields}

The main feature of the stochastic random fields arising as a
result of turbulent processes is that they are correlated. The
central object which describes their properties is the correlation
function $\xi(r)$ (or where more suitable - the structure function
$d(r)$).  Related and often less limited descriptor is the power
spectrum $P(k)$, which is a Fourier transform of the correlation
(structure) functions.

We shall restrict our discussion to the case of
the turbulence that is statistically homogeneous and
isotropic in three dimensional position space \cite{MY75},
\footnote{In the presence of magnetic field MHD turbulence becomes
axisymmetric in the system of reference related to the {\it local}
direction of magnetic field.
The Goldreich \& Sridhar \cite{GoldSrid95} model of incompressible turbulence
prescribes the Kolmogorov scaling of mixing motions perpendicular to magnetic
field lines with $D(r) \propto r^{2/3}$ and a different scaling
along the magnetic field.  Further research in \cite{LG01,CL02a}
has shown that the basic features of the Goldreich-Sridhar turbulence carry 
over for Alfvenic perturbations to the compressible regime.
Observations, however, are usually unable to identify the
local orientation of magnetic field and deal with
the magnetic field projection integrated over the line of sight.
As the result the locally defined perpendicular and parallel directions
are mixed together in the process of observations \cite{CLV02a}.
There is some residual anisotropy, but this anisotropy is scale-independent
and is determined by the rate of the meandering of the large scale
field.  So from the observational point of view
the picture of the isotropic turbulence remains to
some extend applicable. The spectra of intensity fluctuations
obtained from the Goldreich-Sridhar turbulence and the isotropic
turbulence are similar.
}
in which case correlation functions depend only on separation distance
$r$ between two positions in space.
In particular, for the density field $\rho({\bf x})$ we
define the correlation function
\begin{equation}
\xi(r)=\xi({\bf r}) = \langle \rho ({\bf x}) \rho ({\bf x}+{\bf r}) \rangle~~
~.
\label{xifirst}
\end{equation}
or, alternatively, the structure function
\begin{equation}
d(r)=
\langle(\rho({\bf x}+{\bf r})-\rho({\bf x}))^2 \rangle~~~.
\label{2}
\end{equation}

Statistical descriptors can often be assumed to have power-law
dependence on scale $\propto r^{-\gamma}$, where $\gamma$ can be
both positive and negative. The correlation function $\xi(r)$ is an
appropriate choice when stochastic inhomogeneities have more power
on small scales, which corresponds to $\gamma > 0$, while when the
power is concentrated on large scales and $\gamma <0$, the
structure function should be used. With this substitution in mind
both cases can be treated similarly in the most interesting regime
of small $r$.

For power law correlations the power spectrum $P({\bf k})=\int d^N
{\bf r} e^{i{\bf k}{\bf r}} \xi(\bf r)$ also has power law form
$P(k) \propto k^n$, where $n=\gamma-N$. It is important to keep in
mind that the relation between spectral index $n$ and correlation
scaling $\gamma$ involves the dimensionality of space $N$ and is
different for three dimensional fields and two dimensional maps.

To be exact, in the case of density it is more appropriate to
assume a power law scaling for the density fluctuations around the mean
$\delta\rho=\rho-\bar\rho$ rather than for the density field itself, so that
\begin{equation}
\xi(r)=\bar \rho^2 \left[1+\left(r/r_0\right)^{-\gamma}\right] \; .
\label{eq:xirho}
\end{equation}
For example,  in Kolomogorov turbulence the medium is incompressible, however
the passive tracers of the flow, which we may observe as emitters
develop passive scalar density inhomogeneities which scale as
$r^{2/3}$, i.e $\gamma=-2/3$ in our notations.

Whether the medium is incompressible or not, the stochastic motions are the
essence of turbulence.  An isotropic random velocity field ${\bf u}({\bf x})$
is fully described by
the structure tensor $\langle \Delta u_i \Delta u_j \rangle$,
which can be expressed via longitudinal
$D_{LL}$ and transverse $D_{NN}$ components \cite{MY75}
\begin{equation}
\langle \Delta u_i \Delta u_j \rangle = \left( D_{LL}(r)-D_{NN}(r) \right) {r_i r_j \over r^2}
+D_{NN}(r) \delta_{ik}~~~,
\label{struc}
\end{equation}
where $\delta_{ik}$ equals 1 for $i=k$ and zero otherwise.
We need only projection of the velocity structure function onto the
line of sight, which we identify with $z$ direction
$D_z({\bf r}) \equiv \langle \Delta u_i \Delta u_j \rangle \hat z_i \hat z_j$.
For the power-law regime
\begin{equation}
D_z({\bf r}) \sim C r^{m} ~~~
\footnote{There is a residual dependence of $D_z$ on the angle
$\cos\theta={\bf r}\cdot \hat z$ which differs for solenoidal and potential
flows, see \cite{LP00,LP04}. VCA formalism is shown to be insensitive
to it, so an important problem of separating
solenoidal and potential components of the velocity has to be
addressed by combination of techniques, possibly by combining
VCA and velocity centroids \cite{LE03}.}.
\end{equation}
In Kolmogorov turbulence velocity scales the same way, $m=2/3$,
as the density of the passive scalar.  Thus
the Kolmogorov spectrum index is $-11/3$. In turbulence literature
the energy spectrum $E(k)=4\pi k^2 P(k)$ is usually used.
In this notation the Kolmogorov spectrum
is  $E(k)\sim k^{-5/3}$, often referred to as $-5/3$ law.

\subsection{Statistical description in Position-Position-Velocity space}

One does not observe the gas distribution in the real space
galactic coordinates $xyz$ where the 3D vector ${\bf x}$ is
defined. Rather, intensity of the emission in a given spectral
line is defined in Position-Position-Velocity (PPV) cubes towards
some direction on the sky and at a given line-of-sight velocity
$v$ \footnote{All velocities are the line-of-sight velocities, we
omit any special notation to denote z-component.}. In the plane
parallel approximation the direction on the sky is identified with
$xy$ plane where the 2D  spatial vector ${\bf X}$ is defined, so
that the coordinates of PPV cubes available through observations
are $({\bf X},v)$. The relation between the real space and PPV
descriptions is defined by a map $({\bf X},z) \to ({\bf X},v)$.

The central object for our study is a turbulent cloud in PPV
coordinates. The number of particles per unit volume
$\mathrm{d}{\bf X} \mathrm{d}v$ is given by the PPV density
$\rho_s({\bf X},v)$, which statistical properties depend on the
density of gas in real galactic coordinates, but also on velocity
distribution of gas particles. Henceforth, we use the subscript
$s$ to distinguish the quantities in $({\bf X}, v)$ coordinates
from those in $({\bf X}, z)$
 coordinates.
We shall always assume 2D statistical homogeneity and isotropy
of $\rho_s({\bf X},v)$ in ${\bf X}$-direction over the image of a cloud.
 However, homogeneity along the velocity direction can only be assumed after
additional considerations, if at all.
Naturally, there is no symmetry between $v$ and ${\bf X}$.

In \cite{LP04} we discussed in detail the derivation of the correlation
function in PPV space
\begin{equation}
\xi_s(R,v_1,v_2)=\langle\rho_s({\bf R}_1,v_1)\rho_s({\bf R}_2,v_2)\rangle,
~~~~R=\left|{\bf R}_1-{\bf R}_2\right|
\end{equation}
in a general case of matter concentrated in a finite cloud and
possibly subjected to coherent non-random flow, for example
rotation. The main result is that for small scales the
complications arising from the finiteness of the matter
distribution or its non-random motion can be neglected and
\begin{equation}
\xi_s({\bf R},v) \sim \int_{-\infty}^\infty {\mathrm d}z \;
\frac{\xi({\bf r})}{[D_z({\bf r})+2\beta]^{1/2}}
\exp\left[-\frac{v^2}{2 (D_z({\bf r})+2\beta)}\right], ~~~
v=v_1-v_2~, \label{ksicloud}
\end{equation}
where $\beta=k_B T/m$ is related to the thermal velocity of atoms.
How small should the separation $R$ be ? In case of a cloud of
size $S$, obviously we need $R \ll S$ to neglect boundary effects.
In case of the coherent flows, it is the magnitude of the shear
introduced by the flow relative to the magnitude of the stochastic
motions that is important.
Exact criterion is $R \ll \left[f^2 C\right]^\frac{1}{2-m}$
\cite{LP00}, if we parameterize the line-of-sight variation of the
coherent velocity as $v_{coh}=f^{-1} z$. In case of our Galaxy disk
rotation, $f^{-1} \approx 14$ km/s/kpc, while relative turbulent
motions reach $30$ km/s at few tens of parsecs separation. In
\cite{LP00} we have concluded that for scales less than $100$ pc
or perhaps even a bit more, the rotation of the disk is not
important. This argument, on the other side, demonstrates that
HI velocity is a poor indicator of the real line-of-sight position
in PPV cubes.

The equation (\ref{ksicloud}) is an important and rather universal
result. However, it contains several approximations which entered
the derivation and have to be kept in mind:
\begin{itemize}
\item  We have assumed that the underlying turbulent velocity
obeys Gaussian statistics. No such assumptions have been made for
the density, however.
\item Perhaps most importantly, we have
assumed that the velocity and density are uncorrelated. It is
definitely true if they are taken at the same point due to vector
nature of velocity field, but needs not be accurate at finite
separations. An example of a process which may produce
correlations is the self-gravity of the gas, which will point
some velocity vectors towards overdensities. This assumption was
tested numerically in \cite{LPVSP01,ELPC03}. Correlations have to
be quite high to change our results, and that level of correlation
has not been observed.
\item At the final stage, we assume
homogeneity along the velocity dimension as well as in two spatial
directions on the sky. For this to be applicable in practice, one
may have to subtract the average spectral line profile and work
with fluctuations of intensity.
\item The formalism, leading to
the equation (\ref{ksicloud}) has to be extended if the
temperature of the gas varies from point to point and the
turbulence is near subsonic.
\end{itemize}

The correlation in PPV space reflects, as expected, both
underlying inhomogeneities of the matter through $\xi(r)$ and its velocity
through $D_z(r)$.  Important feature of the PPV space is that
$\rho_s({\bf X},v)$
exhibits fluctuations even if the flow is incompressible
and no density fluctuations are present.
Indeed, when one substitutes the expanded expression (\ref{eq:xirho}),
$\xi({\bf r}) =  \bar \rho^2 + \bar\rho^2(r_0/r)^\gamma$, into
eq~(\ref{ksicloud}), both terms give rise to non-trivial
contributions to $\xi_s({\bf R},v)$.

\section{From the line intensity fluctuations
in velocity channels to the characteristics of a turbulent cascade}
Intensity $I_v({\bf X})$ in the spectral line measured at the velocity $v$ in
the direction ${\bf X}$ is obtained by integrating
the standard equation of radiative transfer (Spitzer 1978)
\begin{equation}
dI_{\nu}=-g_{\nu} I_{\nu} dz+j_{\nu}dz
\label{transer1}
\end{equation}
along the line of sight.  In the  case of self-adsorbing
emission in spectral lines the coefficients are proportional
to first power of density:
\begin{eqnarray}
g_{\nu}(z)&=&\alpha(z) \rho(z) \phi_v(z)~~~, \nonumber \\
j_{\nu}(z)&=&\epsilon \rho(z) \phi_v(z)~~~,
\end{eqnarray}
where $\phi_v(z)$ describes the velocity distribution of atoms at the
position $z$ along the line of sight, and all quantities
have implicit dependence on the sky direction ${\bf X}$.
Obvious relation $\int dz \rho(z) \phi_v(z) = \rho_s({\bf X,v})$
links us to the formalism of PPV density.

A solution of the radiative transfer equation
if no external illumination is present
and absorption coefficient $\alpha$ can be taken as constant (the
latter is the essence of the Sobolev approximation) can be written
in a compact form \cite{LP04}
\begin{equation}
I_v({\bf X})=
\frac{\epsilon}{\alpha}\left[1-{\mathrm e}^{-\alpha \rho_s({\bf X},v)}\right]~~~.
\label{simplified}
\end{equation}
In the case of vanishing absorption, the intensity is given by
the linear term in the expansion of the exponent in eq.~(\ref{simplified})
\begin{equation}
I_v({\bf X})=\epsilon \rho_s({\bf X},v)
\label{eq:thinI}
\end{equation}
and reflects the PPV density of the emitters.
If, however, the absorption is strong, the intensity of the emission
is saturated at the value $\epsilon/\alpha$ wherever
$\rho_s({\bf X},v) \gg 1/\alpha$. Identification of the low contrast residual
fluctuations may be difficult in practice.

In observations we register the intensity $I_C({\bf X})$,
integrated over velocity channel
\begin{equation}
I_C({\bf X}) = \int_{-\infty}^\infty {\mathrm d} v W(v-v^*) I_v({\bf X})
\end{equation}
which width and shape is described by the window function $W(v)$.
The properties of the window function are restricted, first of all, by
the experimental setup.
For instance for $CO$ lines studies \cite{FPHPPB98, SBHOZ98}
integration is typically being performed over the whole line, $W(v)=1$.
On the other hand, measurements in 21 cm HI line are performed in
velocity slices of PPV data cube (channel maps), which
corresponds to $W(v-v^*)$ strongly peaked at a particular velocity $v^*$.

When velocity channels are narrow
we have an opportunity to synthetically increase
the width of the channels by combining the data from the adjacent ones.
Within the VCA technique we point out that by  varying the width
of synthetic velocity channels one can separate statistics of turbulent
velocities and density inhomogeneities of emitting medium.
The minimal width of the velocity channel is determined by the resolution
of an instrument. However, thermal motions of the gas lead to the
smearing of intensity fluctuations similar in effect to finite resolution.
Qualitatively, one can think of the minimal width
of the channels available for VCA as given by
a convolution of instrumental and
thermal effects.

Main statistical quantity that we measure from two dimensional
intensity maps $I({\bf X})$ is the structure (correlation) function
\begin{eqnarray}
{\cal D}({\bf R}) &\equiv&
\left\langle \left[I_C({\bf X}_1)- I_C({\bf X}_2)\right]^2\right\rangle,~~~
{\bf R} = {\bf X}_1 - {\bf X}_2 ~~~, \\
\Xi({\bf R}) &\equiv&
\left\langle I_C({\bf X}_1) I_C({\bf X}_2)\right\rangle,~~~
\label{dr_emiss}
\end{eqnarray}
(or, alternatively, 2D power spectrum). Here I shall proceed to describe
how underlying three dimensional statistics are reflected in this
function.

\subsection{Optically thin lines}
From the point of view of VCA approach, the most informative data
comes from optically thin lines, such as, for example HI 21 cm.
In the limit when the absorption can be neglected line intensity
contains direct information about PPV density of emitters (\ref{eq:thinI})
and the correlation function is simply
\begin{eqnarray}
{\cal D}({\bf R}) &=& \epsilon^2 \int\, dv_1 W(v_1) \int\,dv_2 W(v_2)
\left[ d_s({\bf R},v_1,v_2) - d_s(0,v_1,v_2) \right] ~~,\\
\label{eq:optthin}
\Xi({\bf R}) &=& \epsilon^2 \int {\mathrm d} v_1 W(v_1) \int {\mathrm d} v_2
W(v_2) \xi_s({\bf R}, v_1,v_2)~~.
\end{eqnarray}

Let us assume that after subtracting the mean line profile, the
fluctuations of PPV density are statistically homogeneous not only
in ${\bf X}$, but in velocity direction as well\footnote{For an
infinite emitting medium homogeneous turbulence produces a
homogeneous image in the velocity space. For a finite emitting cloud
the PPV image is approximately homogeneous over velocity
separations much less than the Doppler line-width. If we use the
structure function, we do not need to explicitly worry about
subtracting the mean line profile, since its integrand does not
depend on it.}. We shall write out in detail the correlation
function as leading to more compact expressions, but present the
final results in terms of more appropriate ${\cal D}(R)$.  From
the eq.~(\ref{ksicloud}), in the expanded form,
\begin{equation}
\Xi(R) \propto \epsilon^2
\int_{-\infty}^\infty {\mathrm d}z \;
\xi(r) \;
D_z(r)^{-1/2} \int {\mathrm d}v \; W_e^2(v) \;
\exp\left[-\frac{v^2}{2 D_z(r)}\right] ~~,
\label{AppD:xiR}
\end{equation}
where $v$ is the difference in velocities between two emitters, $v=v_1-v_2$,
$v_+=v_1+v_2$ and the effective channel window $W^2_e(v)$ is
a square of the experimental window convolved with the gaussian that
describes the thermal velocities
\begin{displaymath}
W_e^2(v) = \frac{1}{\sqrt{4 \pi \beta}} \int dy
e^{-\frac{(v-y)^2}{4 \beta}}
\int {\mathrm d}v_+\; W_C(v_+-y) W_C(v_++y) ~~~.
\end{displaymath}
For convenience let us parameterize the effective window with a
gaussian shape with the width $\Delta V$, $W_e^2=\Delta V^{-1}
\exp[-v^2/(2 \Delta V^2)]$. The exact shape, of course, depends on
the properties of the instrument.

Equation (\ref{AppD:xiR}) provides the main theoretical foundation for
VCA of optically thin lines. It describes two fundamental regimes:
\begin{description}
\item[thick channel]
The effective channel width $\Delta V$ is larger than the
turbulent velocities at the scale $R$, $\Delta V \gg D_z(R)^{1/2}$.
In this case we collect inside our channel essentially all emitters,
however scattered along the velocity axis they are by the turbulence.
Essentially, $W_e^2 \approx 1$ and
\begin{equation}
\Xi(R) \propto \int_{-\infty}^\infty {\mathrm d}z \; \xi(r) ~~~.
\label{eq:xithick}
\end{equation}
Information about the turbulent velocities is erased, but one can recover
information about density inhomogeneities.
\item[thin channel],
$\Delta V \ll D_z(R)^{1/2}$. Turbulent velocity differences are important,
many emitter pairs close along the line of sight acquire disparate velocities
and do not contribute to the correlation of sources restricted to
the narrow velocity channel.
The amplitude of correlation decreases, but the most important outcome
is that this decrease is scale dependent and we have a change in scaling
law which reflects the underlying velocity statistics. Indeed, in
this case
\begin{equation}
\Xi(R) \propto  W_e^2(0) \int_{-\infty}^\infty {\mathrm d}z \;
\xi(r) \; D_z(r)^{-1/2} ~~.
\label{eq:xithin}
\end{equation}

\end{description}

Interpretation of thick channel data contains no surprises, reflecting
projected density inhomogeneities of the gas.
However if measurements are done in a sufficiently thin channel (and the
criterion depends on scale between two lines of sight), the fluctuations of
intensity may come from two sources. Recall that $\xi(r) \propto
1+ (r/r_0)^{-\gamma}$, therefore
\begin{equation}
\Xi(R) \propto  \int_{-\infty}^\infty {\mathrm d}z \;
D_z(r)^{-1/2} +
\int_{-\infty}^\infty {\mathrm d}z \;
(r/r_0)^{-\gamma} \; D_z(r)^{-1/2} \propto 1+ \Xi_v(R) + \Xi_\rho(R) ~~.
\label{eq:xisplit}
\end{equation}
The first term describes the intensity fluctuations arising just
from turbulent motions of emitters\footnote{this term contains constant
which corresponds just to mean intensity in the map. It has to be regularized
out to make sense of the integral, which is done automatically if structure
functions are used.}.  The second is the velocity modified scaling of
the underlying density fluctuations.
In Table \ref{tab:a} we summarize the predicted scaling for the optically
thin emission intensity in the channel maps.
\begin{table}
\begin{tabular}{lllll}
\hline
  & \tablehead{1}{r}{b}{${\cal D}_v(R)$}
  & \tablehead{1}{r}{b}{${\cal D}_\rho(R)$}
  & \tablehead{1}{r}{b}{$P_v(K)$}
  & \tablehead{1}{r}{b}{$P_\rho(K)$} \\
\hline
Thin channel, $\Delta_V^2 \ll C R^m $ & $\propto R^{1-m/2}$ &
$\propto R^{1-\gamma-m/2}$ & $ \propto K^{-3+m/2}$ & $ \propto K^{n+m/2}$\\
Thick channel, $\Delta_V^2 \gg C R^m $ & 0& $\propto R^{1-\gamma}$&0& $K^n $\\
\hline
\end{tabular}
\caption{Scaling of emission intensity in optically thin lines}
\label{tab:a}
\end{table}
giving both the structure functions, and the
the correspondent two dimensional
power spectra $P_v(K) \propto \int d{\bf R}\; \Xi_v(R)\; e^{i {\bf K R}}$ and
$P_\rho(K) \propto \int d{\bf R}\; \Xi_\rho(R) \; e^{i {\bf K R}}$,

Thus, we are in a position when velocity information can be
obtained from thin channel data,
while increase of channel thickness will determine the density
distribution. If the density scaling is steep, $n < -3$, as is the case
of emitters, behaving as passive scalars
in incompressible Kolmogorov cascade, thin channel measurements
are velocity dominated and exhibit spectra $\propto K^{-3+m/2} \sim K^{-2.7}$
if the stochastic velocities are Kolmogorov, $m=2/3$.

It is a question whether one can obtain thin channels in a
particular observations. The criterion is scale dependent, for
sufficiently small separation between the lines of sight, the
effective width of the channel is always large, but the needed
larger separations may be more difficult to measure. Thermal
effects also increase $\Delta V$. Thus, observing in thin channels
may require special design and be achievable only for cold gas.
However, as we have shown in \cite{LP00}, the HI observations, in
particular \cite{Green93,SL01}, are, effectively, the thin channel
measurements. Thus we may interpret the observed slope of $-2.7$
as a signature of the Kolmogorov stochastic velocities, but we cannot
exclude an alternative that the underlying density inhomogeneities
have enhanced small scale power and $n \approx -3$.

\subsection{Effects of self-absorption on spectral line statistics}

Relation to the underlying turbulence of the intensity in spectral lines
that exhibit self-absorption is much more involved \cite{LP04}.
If the absorption is strong, the intensity of the emission (\ref{simplified})
is saturated at the value $\epsilon/\alpha$ wherever
$\rho_s({\bf X},v) \gg 1/\alpha$. Identification of the low contrast residual
fluctuations may be difficult in practice.

However the study of such lines is crucial, in particular for understanding
molecular clouds. To make a progress, we look for the regimes when intensity
spectra can still exhibit scaling response to the power-law turbulent cascade.
In \cite{LP04} we have found that in the case of weak but not zero absorption,
the universal scaling solution ${\cal D}(R) \propto R$ (this corresponds
to $K^{-3}$ for the power spectrum) arise over the range
of intermediate scales.

Let us return to the structure function ${\cal D}(R)$.
One can always consider sufficiently small scales such
$\alpha^2 \left[d_s({\bf R},v) - d_s(0,v)\right] \ll 1$.
If this holds, we argue in \cite{LP04} that
\begin{equation}
{\cal D}(R) \propto
\int\, dv \; W_e^2(v)\; e^{-\frac{\alpha^2}{2}{d}_s(0,v)} \;
\left[d_s(R,v) - d_s(0,v)\right] ~,
\label{dmainapprox}
\end{equation}
provides a good (although not fully rigorous) approximation to intensity
structure function. Although this expression looks similar to
eq~(\ref{eq:optthin}) for the optically thin case, the conditions
for its validity are more relaxed than a limit $\alpha \to 0$.

The most important effect induced by absorption is retained. It is
an additional exponential down-weighting of the contribution from
the points with large velocity separation $v$ in a manner which
itself depends on the turbulence statistics. This has an effect of
the stochastic velocities having an impact even on the fully
integrated over the frequency lines.

\begin{figure}
  \includegraphics[height=.3\textheight]{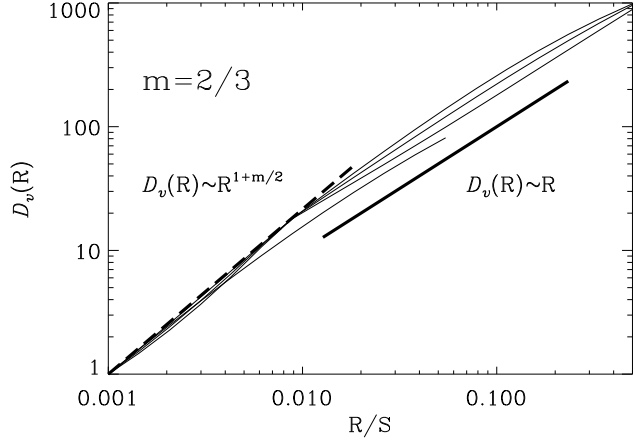}
  \caption{Structure function of intensity fluctuations
    from Kolmogorov turbulent motion in case of weak self-absorption in the
spectral line.}
  \label{fig:a}
\end{figure}

Here we limit ourselves to the short discussion of this effect,
considering density inhomogeneities to be subdominant in PPV statistics.
For more general issues see \cite{LP04}.
Numerical calculations of the ${\cal D}(R)$ for Kolmogorov underlying
velocity given in Figure~\ref{fig:a} show that at large scales $R$
the effects of absorption become
highly nonlinear (indeed, the approximation (\ref{dmainapprox})
breaks down), correlations are erased and the structure function
flattens. At shorter scales, where one may perhaps expect the regime
of absorption behaving like like a thin velocity channel, the
structure function exhibits the $R^1$ slope instead, which we found
to be independent on underlaying velocity scaling
\footnote{The asymptotic analysis in \cite{LP04} shows how
this behavior arises}. At even shorter scales the absorption
becomes negligible and the ordinary thick channel regime takes over.

What determines the scales of transition ?
We may think of absorption as providing a window with the width
$\Delta v_{ab}$ defined by
\begin{equation}
\alpha^2 d_s(0,\Delta v_{ab}) =1~~~~~~~.
\label{vab}
\end{equation}
Only the matter with relative velocities less than $\Delta v_{ab}$
contribute to observed intensity in a correlated fashion.  Smaller
is the $\Delta v_{ab}$, stronger are the absorption effects. This
critical relative velocity depends on the mean density in PPV
space, the absorption coefficient, and the slope of the PPV
structure function. How important the absorption is for intensity
correlations at a scale $R$ depends on comparison between $\Delta
v_{ab}$ and {\it rms} turbulent velocity at this scale,
$D_z(S)^{1/2} (R/S)^{m/2}$ where $S$ is the size of our cloud. If
$\Delta v_{ab}$ is larger than {\it rms} velocity differences
between emitters, its effect is gradually diminished. This
translates into absorption being less important on small scales
and more important on large scales. A detailed account of different
regimes and transition scales is given in Table~\ref{tab:b}.
\begin{table}[t]
\begin{tabular}{lll}
\hline
    \tablehead{1}{l}{b}{Scale Range}
  & \tablehead{1}{l}{b}{Intensity Scaling}
  & \tablehead{1}{l}{b}{Regime} \\
\hline
$R/S < (\beta/D_z(S))^{1/m}$ & velocity effects erased & (subsonic~regime) \\
$R/S \ll \left(\frac{v_{ab}^2}{D_z(S)}\right)^{1/m}$ &
${\cal D}(R) \propto R^{1+m/2}$ &  (transition to thick slice) \\
$R/S < \left(\frac{v_{ab}^2}{D_z(S)}\right)^{1/m} $&
${\cal D}(R) \propto R^1 $& (intermediate scaling) \\
$\left(\frac{v_{ab}^2}{D_z(S)}\right)^{1/m} <
R/S < \left(\frac{v_{ab}^2}{D_z(S)}\right)^{2/(2-m)}$ &
${\cal D}(R) \propto R^{1-m/2}$ & (thin slice) \\
$\left(\frac{v_{ab}^2}{D_z(S)}\right)^{2/(2-m)} < R/S $
& not a power law & (strong absorption regime) \\
\hline
\end{tabular}
\caption{Scalings of structure functions of intensity fluctuations arising
from velocity fluctuations for the power-law
underlying 3D velocity statistics. In the strong absorption regime
${\cal D}(R)$
does not follow a simple power-law. The regime of thin channel
is not realized for $m \ge 2/3$, including the Kolmogorov case.}
\label{tab:b}
\end{table}

In view of this analysis $CO$ measurements of \cite{SBHOZ98}
which show shallower, $-2.7$, slope instead of $-3$
may be taken to indicate that in molecular clouds
the density inhomogeneities have
enhanced small scale power with $n > -3$ and dominate the velocity
contribution. However to make definitive statement one needs to make
sure that the difference of $0.3$ in the power spectrum can not arise
from the noise in the data. $\delta n=0.3$ or better seems to be
an accuracy we need to have to distinguish between different theoretical
possibilities discussed in this paper.

\section{Summary}
We have reviewed
the formalism of Velocity Channel Analysis for optically thin lines
and its extension to the lines with self-absorption.
We demonstrate that by observing optically thin lines from cold gas
in sufficiently narrow (thin) velocity channels one may recover
the scaling of the stochastic velocities from turbulent cascade,
in particular, Kolmogorov velocities give $K^{-2.7}$ contribution
to the intensity power spectrum. This matches the observational
data in HI, both in our Galaxy and in Small Magellanic cloud.
Synthetically increasing the channel thickness separates out
the underlying density inhomogeneities of the gas. An attempt to apply
this technique to HI lines in SMC \cite{SL01} indeed showed
the expected steepening of the spectrum with the channel thickness.

Effects of self absorption, on the other hand, retain the velocity
signature even for integrated lines. As a result, intensity
fluctuations from velocity tend to show universal scaling of the power
$\propto K^{-3}$ over the range of scales. Observed shallower spectra
$n \approx -2.6$ in CO lines therefore may indicate that real
density spectrum is also shallow $n \approx -2.6$ and dominates the signal.

Progress in interpretation of the data require the knowledge of the power
spectrum index better than with an accuracy of 0.3 and developing
of the accurate theoretical and numerical models to match such accuracy
in different regimes.

\begin{theacknowledgments}
Research of AL is supported by the NSF Grant  AST-0307869 and by
NSF Center for Magnetic Self-Organization in Laboratory and Astrophysical
Plasmas, while research of DP is supported
by the Natural Sciences and Engineering Research Council of Canada.
\end{theacknowledgments}


\begin{thebibliography}{9}

\bibitem{Arm1995}
J.M. Armstrong,  B.J. Rickett \& S.R. Spangler, \emph{Astrophys.
J.}, \textbf{443}, 209, (1995).
\bibitem{Laz99a}  A. Lazarian,
``Statistics of Turbulence from Spectral-Line Data Cubes,'' in
\emph{Plasma Turbulence and Energetic Particles}, ed. M. Ostrowski
and R. Schlickeiser, Cracow (1999), p28.
\bibitem{LPE02}
 A. Lazarian, D. Pogosyan \& A. Esquivel,  in \emph{Seeing through the
dust}, ed. Taylor R., Landecker T. L., Willis A. G., ASP Conf.
Ser. Vol. 276, Astron. Soc. Pac., San Francisco, 2002, p. 182.
\bibitem{Dickman85}
R.L. Dickman,  ``Turbulence in Molecular Clouds,'' in
\emph{Protostars and Planets II}, eds Black~D.C. and Mathews~M.S.,
Tucson:  University of Arizona, , 1985, p150.
\bibitem{Cordes99}
 J. Cordes, ``The Spectrum \& Galactic Distribution of
MicroTurbulence in Diffuse Ionized Gas,'' in \emph{Interstellar
Turbulence},
 eds. Jose Franco \& Alberto Carraminana, (1999), p.33.
\bibitem{Laz99b} A. Lazarian,  ``Turbulence in Atomic Hydrogen,'' in
\emph{Interstellar Turbulence}, eds. Jose Franco \& Alberto
Carraminana (1999), p.95 .
\bibitem{GoldSrid95} P. Goldreich  \& S. Sridhar,
\emph{Astrophys. J.}, \textbf{438}, 763 (1995).
\bibitem{LG01} Lithwick, Y. \& Goldreich, P. 2001,
\emph{Astrophys. J.}, \textbf{562}, 279 (2001).
\bibitem{CL02a} J. Cho \& A. Lazarian,  \emph{Phys. Rev. Lett},
\textbf{88}, 5001, (2002).
\bibitem{CLV02a} J. Cho,  A. Lazarian \& E. Vishniac, in
\emph{Simulations of magnetohydrodynamic turbulence in
astrophysics}, eds. T. Passot \& E. Falgarone (Springer LNP),
2002, p56.
\bibitem{CL03a} J. Cho \& A. Lazarian,  in \emph{Acoustic emission and
scattering by turbulent flows}, ed. M. Rast, Springer LNP, 2003.
\bibitem{CL03b} J. Cho \& A. Lazarian, \emph{Mon. Not. Roy. Astron.  Soc},
\textbf{345}, 325 (2003).
\bibitem{Schli99}  R. Schlickeiser, ``Quasilinear Theory of Cosmic
Ray Transport,'' in \emph{Weak Magnetohydrodynamic Plasma
Turbulence}, ed. M. Ostrowski and R. Schlickeiser, Cracow, 1999,
225.
\bibitem{VOPGS00} E. V\'azquez-Semadeni,   E. C. Ostriker, T. Passot,
C. Gammie  \& J. Stone,  in \emph{Protostars \& Planets IV}, eds.
V. Mannings, A. Boss \& S. Russell, Univ.\ of Arizona Press,
Tucson, 2000, p3 .
\bibitem{NM01} R. Narayan \& M.V. Medvedev,
\emph{Astrophys. J.}, \textbf{562}, L129 (2001).
\bibitem{CLV03} J. Cho, A. Lazarian \& E. T. Vishniac,
\emph{Astrophys. J.}, \textbf{595}, 812 (2003).
\bibitem{MY75}  A.S. Monin,, \& A.M. Yaglom,
``Statistical Fluid Mechanics: Mechanics of Turbulence,'' vol. 2,
The MIT Press (1975)
\bibitem{Dickey95}
J.M. Dickey,  in \emph{The Physics of the Interstellar Medium},
eds, Pfenninger, D. and Bartholdi, P., Springer-Verlag, 1995, p.1.
\bibitem{LP00}  A. Lazarian \& D. Pogosyan,
\emph{Astrophys. J.}, \textbf{537}, 720 (2000).
\bibitem{LP04} A. Lazarian  \& D. Pogosyan,
\emph{Astrophys. J.}, \textbf{616}, 943 (2004).
\bibitem{Green93} D.A. Green,
\emph{Mon. Not. Roy. Astr. Soc}, \textbf{262}, 328 (1993).
\bibitem{DMcSGG01} J.M. Dickey, N.M. McClure-Griffiths, S. Stanimirovic,
Gaensler, B.M., \& Green, A. J., \emph{Astrophys. J.},
\textbf{561}, 264 (2001).
\bibitem{SL01} S. Stanimirovi\'{c}  \& A. Lazarian,
\emph{Astrophys. J. Lett}, \textbf{551}, 53 (2001)
\bibitem{SBHOZ98} J. Stutzki,  F. Bensch, A. Heithausen, V. Ossenkopf,
\& M. Zielinsky,  \emph{Astron. Astrophys}, \textbf{336}, 697,
(1998).
\bibitem{MS96}  A. H. Minter  \& S. R. Spangler,
\emph{astrophys. J.}, \textbf{458}, 194 (1996).
\bibitem{Kolm41} A. Kolmogorov,  \emph{Compt. Rend. Acad. Sci. USSR},
30, 301 (1941).
\bibitem{LE03} A. Lazarian  \&  A. Esquivel,
\emph{Astrophys. J.}, \textbf{592}, L37 (2003).
\bibitem{LPVSP01} A. Lazarian, D. Pogosyan,  E. Vazquez-Semadeni
\& B. Pichardo, \emph{Astrophys. J.}, \textbf{555}, 130 (2001).
\bibitem{ELPC03}
A. Esquivel , A. Lazarian , D. Pogosyan  \& J. Cho , \emph{Mon.
Not. Roy. Astron. Soc}, \textbf{342}, 325 (2003).
\bibitem{FPHPPB98} E. Falgarone,  J.-F. Panis,  A. Heithausen,
J. Perault, J.-L.  Puget \& F. Bensch,  \emph{Astron. Astrophys.},
\textbf{331}, 669 (1998).

\end{thebibliography}
\end{document}